\documentclass[prd,aps,showpacs,twocolumn,preprintnumbers,nofootinbib,amsmath,amssymb,superscriptaddress,longbibliography]{revtex4-2}

\usepackage{systeme}
\usepackage{mathtools}
\usepackage[inline]{enumitem}
\usepackage{graphicx}

\usepackage[T1]{fontenc}
\usepackage[english]{babel}
\usepackage{xcolor}
\usepackage{braket}
\usepackage{nicefrac}
\usepackage{bm}
\usepackage{bbm}
\usepackage{braket}
\usepackage{slashed}
\usepackage[normalem]{ulem}
\usepackage{array}
\newcolumntype{C}{>{$}c<{$}}

\newcommand{\beqn}{\begin{eqnarray}}
\newcommand{\eeqn}{\end{eqnarray}}
\newcommand{\beqs}{\begin{subequations}}
\newcommand{\eeqs}{\end{subequations}\\[-2mm]\noindent}
\newcommand{\eq}[1]{(\ref{#1})}

\newcommand{\bs}{\boldsymbol}

\begin{document}
\bibliographystyle{apsrev4-1}

\title{Casimir boundaries, monopoles, and deconfinement transition in 3+1 dimensional compact electrodynamics}

\author{M. N. Chernodub}
\affiliation{Institut Denis Poisson UMR 7013, Universit\'e de Tours, 37200 Tours, France}
\author{V. A. Goy}
\affiliation{Pacific Quantum Center, Far Eastern Federal University, 690950 Vladivostok, Russia}
\author{A. V. Molochkov}
\affiliation{Pacific Quantum Center, Far Eastern Federal University, 690950 Vladivostok, Russia}
\author{A. S. Tanashkin}
\affiliation{Pacific Quantum Center, Far Eastern Federal University, 690950 Vladivostok, Russia}

\begin{abstract}
Compact U(1) gauge theory in 3+1 dimensions possesses the confining phase, characterized by a linear raise of the potential between particles with opposite electric charges at sufficiently large inter-particle separation. The confinement is generated by condensation of Abelian monopoles at strong gauge coupling. We study the properties of monopoles and deconfining order parameter in zero-temperature theory in the presence of ideally conducting parallel metallic boundaries (plates) usually associated with the Casimir effect. Using first-principle numerical simulations in compact U(1) lattice gauge theory, we show that as the distance between the plates diminishes, the vacuum in between the plates experiences a deconfining transition. The phase diagram in the space of the gauge coupling and the inter-plane distance is obtained. 
\end{abstract}

\date{\today}

\maketitle

\section{Introduction}

Physical objects affect fluctuations of quantum fields and modify dispersion relations of quantum fluctuations in the vacuum around them. This phenomenon is the essence of the Casimir effect~\cite{ref_Casimir}, which predicts that the energy of vacuum fluctuations is modified by the presence of physical bodies~\cite{ref_Bogdag, ref_Milton}. Moreover, the shift in the energy of virtual particles has a fundamental physical consequence as the Casimir effect results in a tiny force between neutral objects~\cite{Casimir_1947kzi} that can be detected experimentally~ \cite{link_experiment_1,link_experiment_2,link_experiment_3}. The Casimir effect is one of the demonstrations of the importance of vacuum fluctuations and the physical significance of the mysterious vacuum energy.

Most straightforwardly, the Casimir effect reveals itself in non-interacting field theories. However, even in the absence of interactions, the shift in vacuum energy sets a complex analytical problem apart from a few simplest geometries of physical bodies. Therefore, the Casimir effect is usually studied using, for example, the analytical proximity force approximation~\cite{ref_proximity}, or utilizing numerical tools~\cite{Johnson_2010ug} that include the world line approaches~\cite{Gies_2006cq} and first-principle methods of lattice gauge theory~\cite{ref_Oleg_1,ref_Oleg_2,ref_Oleg_3,ref_Oleg_4,ref_paper_1,ref_paper_2,ref_paper_3,ref_paper_4}.

In the presence of (self-) interactions of fields, for example, in quantum electrodynamics (QED), the calculations of the Casimir interaction become much more involved. However, in the experimentally relevant cases, the effect of interactions of fundamental fields is very small due to the weakness of the QED coupling constant. The electron-photon coupling affects the Casimir-Polder force in the second-order of the perturbation theory, thus making this contribution undetectable with current experimental techniques~\cite{Bordag_1983zk}.

In strongly coupled field theories, interactions can substantially change the magnitude of the Casimir force and, unexpectedly, modify the vacuum structure of the theories themselves. In theories with dynamical matter fields, the presence of reflective boundaries can affect vacuum condensates and generate new (and modify existing) phase transitions such as the chiral phase transition in the four-fermion effective field theory~\cite{Flachi_2013bc,Tiburzi_2013vza,Chernodub_2016kxh}. Interactions can also change the sign of the Casimir--Polder force in fermionic systems with condensates~\cite{Flachi_2017cdo}, as well as in the $\mathbb{C}P^{N-1}$ model~\cite{Flachi_2017xat} (see also \cite{Betti_2017zcm}).

The Casimir geometry can also affect non-perturbative phenomena associated with the gauge (vector) fields. For example, vacua of compact Abelian U(1) gauge theory and non-Abelian SU(2) gauge theory in two spatial dimensions lose the confining property at sufficiently small separations between, respectively, perfectly metallic and chromometallic  plates~\cite{ref_paper_3,ref_paper_4}. In compact Abelian theory, the deconfining transition is associated with the binding transition of the Abelian monopoles, which emerges in the monopoles plasma due to the presence of the boundaries~\cite{ref_paper_3}. 

We study the effect produced by two closely-spaced, perfectly-conducting boundaries on the phase structure of 3+1 dimensional compact U(1) gauge theory. The study may be relevant to the MIT bag model, which treats the hadron as a (spherical) deconfinement region separated from the confining exterior by a (reflective) wall~\cite{MIT_1,MIT_2}. Thermodynamically, the wall is supported from the collapse by a Casimir pressure of this finite-volume system~\cite{MIT_3,MIT_4,Horowitz_2021dmr}. We also investigate the simplest statistical properties of the Abelian monopoles in the space between the plates since these topological objects are responsible for the confinement of charge in compact U(1) gauge theory.

The structure of this paper is as follows. In Section~\ref{sec_model} we describe the lattice model, the monopoles, and the definition of the perfectly metallic Casimir boundary conditions. Then, in Section~\ref{sec_monopoles} we show how the Casimir plates affect the monopoles. Next, we use the monopole properties to determine the phase diagram of the model. The deconfinement order parameter in the inter-plate space are discussed in Section~\ref{sec_confinement}. The last section is devoted to our conclusions.

\section{Compact electrodynamics on the lattice, Casimir plates and monopoles}
\label{sec_model}

\subsection{The model}

We study a 3+1 dimensional compact U(1) gauge theory in lattice regularization suitable for numerical simulations. The calculations in thermal equilibrium are performed, after Wick rotation, in four Euclidean spacetime dimensions. Below we briefly discuss the model, topological defects, and the Casimir boundary conditions, following the discussion in 2+1 dimensional case~\cite{ref_paper_1,ref_paper_2} closely.

The compact U(1) gauge model describes the dynamics of the lattice gauge (photon) field $\theta_{x,\mu} \in [-\pi,+\pi)$ which is defined on the elementary links $l=\{x,\mu\}$ set by its starting point $x$ and the direction $\mu$. In the continuum limit, $a\to 0$, the lattice field $\theta_{x\mu} = a A_{\mu}(x)$ is related to the continuum gauge field $A_\mu(x)$ and the lattice spacing~$a$. The lattice analogue of the field-strength tensor $F_{\mu\nu} \equiv \partial_\mu A_\nu - \partial_\nu A_\mu$ is played by the plaquette angle,
\beqn
\theta_{P_{x,\mu\nu}} = \theta_{x,\mu} + \theta_{x+\hat\mu,\nu} - \theta_{x+\hat\nu,\mu} - \theta_{x,\nu}\,,
\label{eq_theta:P}
\eeqn
constructed from the link fields (link angles) $\theta_{x,\mu}$. Each plaquette $P \equiv P_{x,\mu\nu}$ is set by the position~$x$ of one of its corners and by two vectors of the plaquette plane, $\mu < \nu$, with the axes labeled by the indices $\mu,\nu = 1, \dots, 4$. The indices $\mu=1,2,3$ correspond to spatial directions while $\mu = 4$ marks the imaginary Euclidean time.

In the continuum limit, the plaquette angle~\eq{eq_theta:P} reduces to its continuum analogue $\theta_{P_{x,\mu\nu}} = a^2 F_{\mu\nu}(x) + O(a^4)$ for small (perturbative) fluctuations of the photon field. In addition to the perturbative fluctuations, the model also possesses the topological configurations of the gauge fields, the Abelian monopoles, which correspond to large variations of the lattice gauge field $\theta_{x,\mu} \sim 1$. These configurations are singular in the continuum limit.

The action of the lattice model,
\beqn
S[\theta] = \beta \sum_P \left(1 - \cos \theta_P \right)\,,
\label{eq_S}
\eeqn
is given by the sum over all elementary lattice plaquettes $P$. For configurations without monopoles, the lattice action~\eq{eq_S} becomes the standard photon action if one associates the lattice coupling constant $\beta = 4/e^2$ with the electric charge $e$. In the presence of the Abelian monopole singularities, the continuum action becomes more complicated as it includes singular Dirac sheets attached to the worldlines of the Abelian monopoles. The continuum formulation of the compact QED has been briefly discussed in one of our previous papers~\cite{ref_paper_2}.

The model~\eq{eq_S} is also called the ``compact'' model because the Abelian gauge group of the theory corresponds to a compact manifold, $S^1$. The action is invariant under discrete shifts of the plaquette variable, $\theta_P \to \theta_P + 2 \pi n_P$ with an integer $n_P \in {\mathrm Z}$. This invariance implies that two lattice field strengths $\theta_P$ and $\theta_P' = \theta_P + 2 \pi n$ with $n \in {\mathbb Z}$ are physically equivalent to each other, thus reducing the physical gauge group to a circle, ${\mathbb R}/{\mathbb Z} \sim S^1$. The same symmetry is also applied to the gauge field itself, $\theta_{x,\mu} \to \theta_{x,\mu} + 2 \pi k_{x,\mu} $ with $k_{x,\mu} \in {\mathrm Z}$.

\subsection{Magnetic monopoles and electric confinement}

The compactness of the model naturally leads to the appearance of singular configurations of the gauge field, the Abelian monopoles. In the continuum limit, the mentioned $2 \pi$ shifts, which leave invariant the lattice action~\eq{eq_S}, correspond to the physically unobservable displacements of the singular Dirac sheets (i.e., the worldlines of the Dirac strings attached to the Abelian monopoles). The end-points of the open Dirac strings correspond to the trajectories of the Abelian monopoles, which are physical, gauge-invariant topological defects. 

The monopoles are particle-like objects in the (3+1)-dimensional compact electrodynamics.  On the lattice, the monopole current $j_{x,\mu}$ can be determined via a finite-difference divergence of the physical part 
\beqn
{\bar \theta}_P = \theta_P + 2 \pi k_P \in [-\pi,\pi)\,, \qquad k_P \in {\mathrm Z}\,,
\label{eq_bar:theta}
\eeqn
of the lattice field-strength tensor $\theta_P$. The monopole trajectory corresponds to a collection of three-dimensional cubes $C_{x,\mu}$ which contain a nonzero magnetic charge, $j_{x,\mu} \neq 0$:
\beqn
j_{x,\mu} = \frac{1}{2\pi} \sum_{P \in \partial C_{x,\mu}} {\bar \theta}_P \in {\mathbb Z}\,.
\label{eq_j:lattice}
\eeqn
Here the sum goes over all elementary sides $P$ of the cube $C_{x,\mu}$, and the index $\mu$ specifies the local direction of the monopole current. The index $\mu$ is normal to the three axes that form the 3d cube $C_{x,\mu}$. For example, if $j_{x,4} \neq 0$, then the corresponding 3-cube is a spatial cube which contains a static segment of the monopole trajectory. 

In the continuum limit, the sum in Eq.~\eq{eq_j:lattice} reduces to the divergence of the magnetic field, thus signaling violation of the Bianchi identities for singular field configurations, $\epsilon^{\mu\nu\alpha\beta} \partial_\nu F_{\alpha\beta} \neq 0$. The monopole trajectory~\eq{eq_j:lattice} forms a closed loop defined at the dual hypercubic lattice (for a review, see Ref.~\cite{Chernodub_1997ay}). 

It is convenient to define the global monopole density,
\beqn
\rho = \frac{1}{{\mathrm{Vol}}_4} \sum_{x,\mu} |j_{x,\mu}|\,,
\label{eq_rho:lattice}
\eeqn
where the sum is performed over the volume Vol${}_4$ of the four-dimensional hypercubic lattice. We will also calculate the density~\eq{eq_rho:lattice} and the related quantities in restricted volumes (in between the Casimir plates).

In lattice gauge theories, the Abelian monopoles have been intensively probed for their possible role in the phenomenon of the charge confinement, which is, presumably, closely related to the color confinement in non-Abelian gauge theories such as QCD~\cite{Mandelstam_1974pi,tHooft_1981bkw,Chernodub_1997ay}. The monopole condensation in an Abelian gauge theory leads to linear confinement of the electric charges because the monopole condensate squeezes the electric flux emanating from the electric charges into a thin electric tube. The tube plays the role of a confining string. Since the string is a linear object with a constant energy density $\sigma$ per string length, the increasing distance $R$ between the particle-antiparticle pair leads to a linearly rising potential $V(R) \simeq  \sigma R$ at large distances. The dimensionful parameter $\sigma$ has the sense of string tension.

This confining mechanism is similar (and dual) to the formation of the Abrikosov vortices in superconductors, where the electrically charged condensate of electron Cooper pairs squeezes the magnetic flux into thin vortices. If a monopole--anti-monopole pair were immersed into the superconductor, it would be confined due to the appearance of the Abrikosov vortex stretched between the constituents of the pair. The mechanism of the charge confinement based on the monopole condensation is often called the dual superconductor mechanism~\cite{Mandelstam_1974pi,tHooft_1981bkw}.

The dual superconductor mechanism was shown to work in the four-dimensional compact electrodynamics, which possesses the straightforward and unambiguous definition~\eq{eq_j:lattice} of the Abelian monopoles~\cite{DeGrand_1980}. The onset of the monopole condensation, related to percolation of monopole trajectories~\cite{Ivanenko_1991wt}, is a well-defined phenomenon in this model. More complicated, non-Abelian Yang-Mills theories were also shown to possess the dual superconductivity phenomenon in low-temperature, confining phase~\cite{Ivanenko_1991wt,Suzuki_1989gp,Stack_1994wm,Bali_1996dm,Chernodub_1996ps,DiGiacomo_1999yas}.

\subsection{Casimir plates for compact gauge fields}

In (3+1) dimensions, the Casimir problem is defined, in general, for three-dimensional physical materials possessing two-dimensional surfaces. If the surfaces are made of an ideal metal, then two tangential (to the surface at each point) components of the electric field and a normal component of the magnetic field vanish. These boundary conditions can be written in a covariant form:
\beqn
\varepsilon^{\mu\nu\lambda\sigma}F_{\nu\lambda}(x) v_{\sigma}(x) = 0\,, \qquad \mu = 1,\dots 4\,,
\label{eq_F:0:cov}
\eeqn
where
\beqn
v_{\mu}(x) = \varepsilon_{\mu\nu\lambda\sigma}\int d^3 \xi \, \frac{\partial {\bar x}^{\nu}}{\partial \xi_1} \frac{\partial {\bar x}^{\lambda}}{\partial \xi_2} \frac{\partial {\bar x}^{\sigma}}{\partial \xi_3} \,
\delta^{(4)}\bigl(x - {\bar x}({\vec\xi}\,)\bigr), \quad
\label{eq_s:munu:gen}
\eeqn
is the dual 3-volume element of the world sheet of the surface. The latter is described by the vector function ${\bar x}^\mu = {\bar x}^\mu(\vec\xi\,)$ parameterized by the three dimensional vector $\vec\xi = (\xi_1,\xi_2,\xi_3)$.

\begin{figure}[h]
	\centering
	\includegraphics[width=55mm]{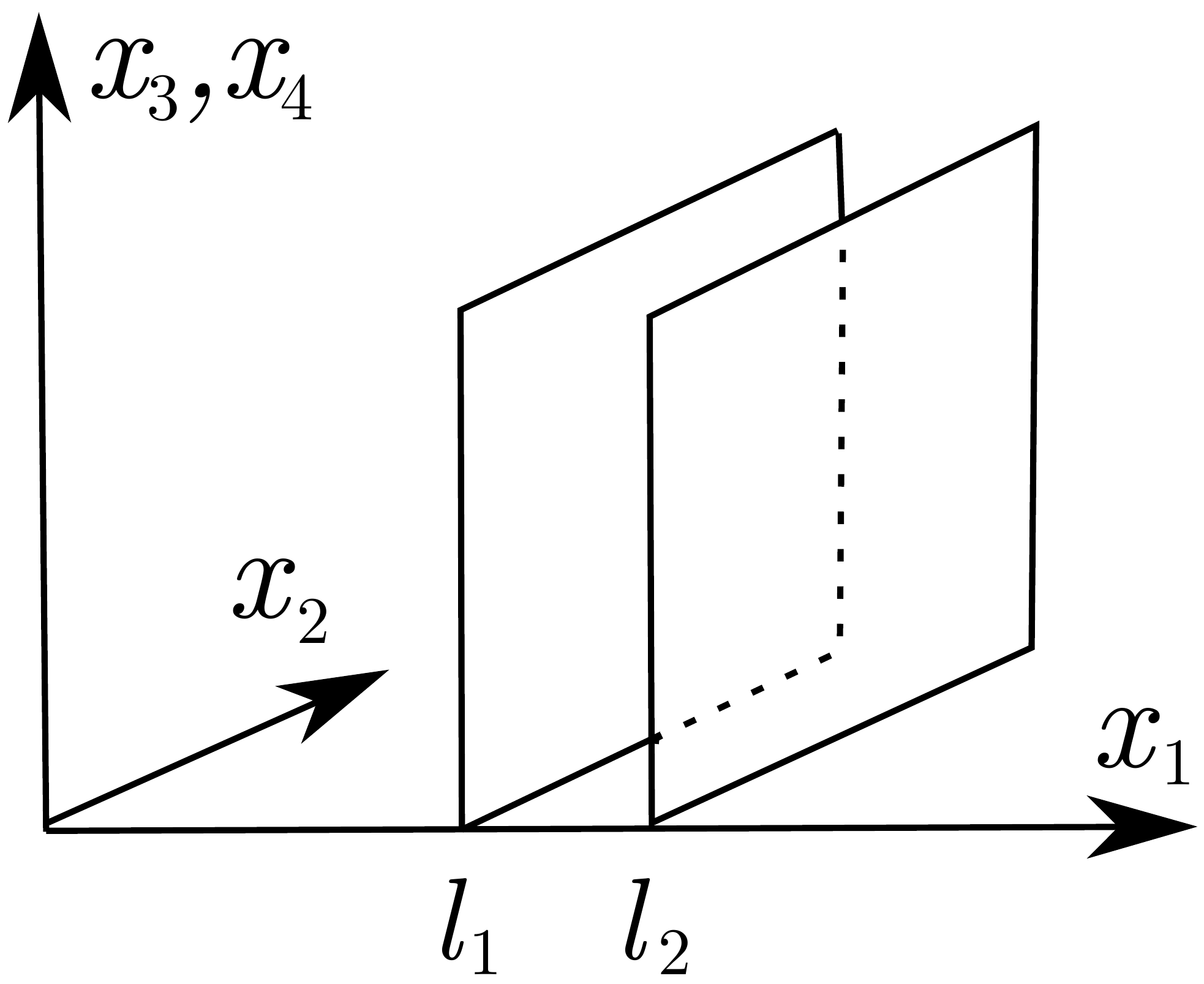}
	\caption{The geometric setup of two parallel Casimir plates.}
	\label{fig:setup}
\end{figure}

In our paper, we consider two static flat plates normal to the $x_1$ axis set at the positions $x_1 = l_1$ and $x_1 = l_2$ as shown in Fig.~\ref{fig:setup}. For each plate, the local volume element of the corresponding world volume~\eq{eq_s:munu:gen} is given by the following formula
\beqn
\nu_{\mu}(x) = \delta_{\mu,1} \delta(x_1 - l_a)\,,  \qquad a = 1,2\,,
\eeqn
where the parameter $a$ labels the surfaces. To derive the above formula, one can use the parameterization of the $a$th surface as follows: ${\bar x}^\mu_a = (l_a,\xi_1,\xi_2,\xi_3)$. Consequently, the covariant conditions~\eq{eq_F:0:cov}, reduce to the three conditions that include  the normal component of the magnetic field and two tangential components of the (Euclidean) electric field, respectively:
\beqn
B_1 & \equiv & F^{23}(x) {\biggl|}_{x_1 = l_a} = 0\,, \\
E_2 & \equiv & F^{24}(x) {\biggl|}_{x_1 = l_a} = 0\,, \\\
E_3 & \equiv & F^{34}(x) {\biggl|}_{x_1 = l_a} = 0\,.
\label{eq_F:Casimir}
\eeqn

The definition of the lattice field strength tensor~\eq{eq_theta:P} and its physical part~\eq{eq_bar:theta}, imply that the lattice Casimir conditions~\eq{eq_F:Casimir} read as follows~\cite{ref_paper_1,ref_paper_2}:
\beqn
\cos\theta_{x,\mu\nu} {\biggl|}_{x_1 = l_a} = 1, \qquad (\mu,\nu) = (23,24,34)\,,
\label{eq_F01:latt:3D}
\eeqn
for all possible $(x_2,x_3,x_4)$ and fixed $x_1 = l_a$  with $a = 1,2$. 

The simplest way to implement the boundary condition~\eq{eq_F01:latt:3D} in the path integral formalism is to add a set of Lagrange multipliers to the standard action~\eq{eq_S}:
\beqn
S_{\varepsilon}[\theta] = \sum_P \beta_P(\varepsilon) \cos \theta_P\,,
\label{eq_S:beta}
\eeqn
To this end, we modify the lattice plaquette couplings $\beta_P \to \beta_P(\varepsilon)$. The inhomogeneous coupling
\beqn
\beta_{P} (\varepsilon) = \beta \bigl[1 + (\varepsilon - 1)\, \delta_{P,{\mathcal V}}\bigr]\,.
\label{eq_beta:P}
\eeqn
is a function of the dielectric permittivity $\varepsilon$ of the Casimir plates.  Here we used the notation $\mathcal V$ to denote the collection of plaquettes $P_{x,\mu\nu}$ belonging to the world volume of the plates. The plates are effectively absent if $\varepsilon = 1$ while in the limit $\varepsilon \to \infty$, the components of the physical lattice field-strength tensor~\eq{eq_bar:theta} vanish at the word-volume of the plates and, consequently, the lattice condition~\eq{eq_F01:latt:3D} is satisfied. In this limit, the plates become perfectly metallic~\cite{ref_Bogdag}. 

We perform our simulations on a $24^4$ lattice corresponding to a zero-temperature model. To generate and update gauge field configurations, we used the Monte-Carlo heatbath algorithm~\cite{Gattringer_2010zz,ACP}. For each point, set by gauge coupling constant $\beta$ and distance between plates $R$, we generated $7.5\times 10^5$ trajectories. The first $10^5$ configurations are omitted to achieve thermalization. 

\section{Monopoles and Casimir plates}
\label{sec_monopoles}

\subsection{Monopole density in the absence of the plates}

It is well known that in the strong coupling region of the theory, $\beta \lesssim 1$, the monopole trajectory forms a dense percolating cluster~\cite{Ivanenko_1991wt}. This cluster corresponds to the monopole condensate, which, according to the dual superconducting scenario, produces the confinement of oppositely charged electric test particles. The property of percolation implies that any of two points in space have a nonzero probability of being connected by a monopole trajectory. Furthermore, the infrared nature of the monopole condensate implies that at infinitely large separation between the points, the nonzero percolation probability stays constant as the distance between the points increases.

In the weak coupling regime, $\beta \gtrsim 1$, the percolation cluster breaks down, thus signaling that the monopole condensate disappears. The confinement property is, consequently, lost. The confining strongly-coupled regime and the weakly coupled deconfining phase are separated by the first-order phase transition~\cite{Arnold_2002jk,Vettorazzo_2003fg}. While we do not study the percolation properties of the monopole cluster in our paper, the transition point can be deduced from the behavior of the much simpler quantity, the monopole density~\eq{eq_rho:lattice}. 

\begin{figure}[!htb]
\begin{center}
            \includegraphics[width=0.45\textwidth]{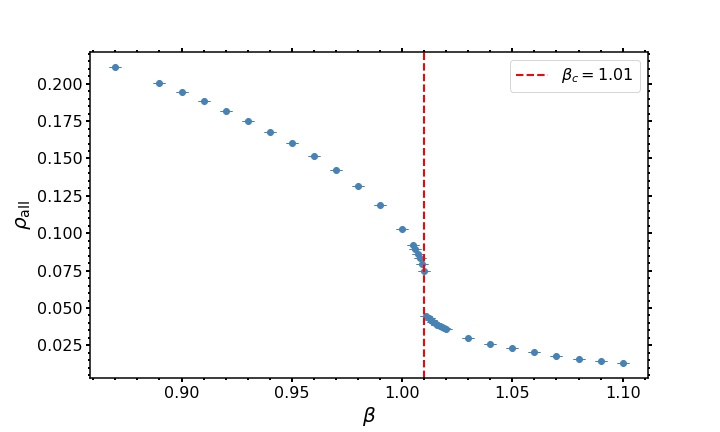} \\ (a) \\[1mm]
            \includegraphics[width=0.45\textwidth]{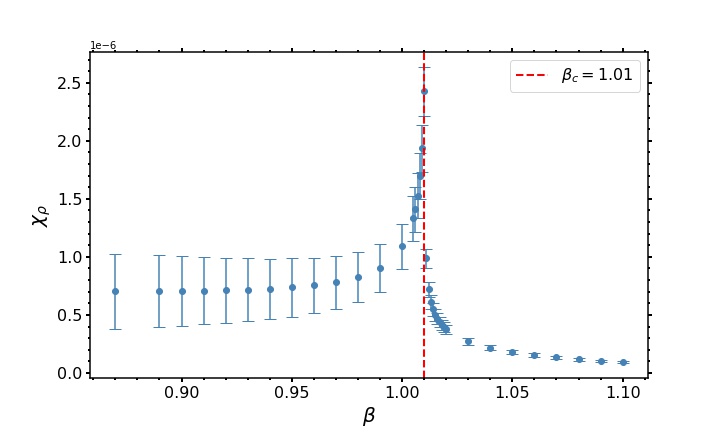}\\ (b) \\
            \includegraphics[width=0.45\textwidth]{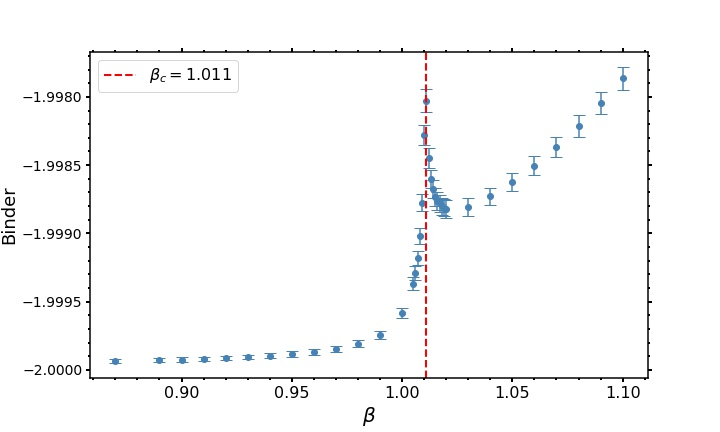}\\ (c)
\end{center}
        \caption{(a) Monopole density $\rho$, (b) its susceptibility and (c) the Binder cummulant~\eq{eq_binder} for ${\cal O} = \rho$ vs lattice coupling $\beta$ in the absence of plates. The vertical line marks the position of the phase transition calculated from these observables.}
    \label{fig:mdsusc_nocas}
\end{figure}

Before performing calculations of Casimir effects, we consider case of the homogeneous lattice in the absence of metallic plates. In Fig.~\ref{fig:mdsusc_nocas}, we show the monopole density~$\rho$, the corresponding susceptibility,
\beqn
\chi_{\rho} = \langle \rho^2 \rangle - \langle \rho \rangle^2\,,
\label{eq_suscept}
\eeqn
and the Binder cummulant (with $\rho = \rho$),
\beqn
    B_{\rho} = \frac{\langle \rho^4 \rangle}{\langle \rho^2 \rangle^2} - 3\,,
    \label{eq_binder}
\eeqn
as the functions of the lattice coupling $\beta$. The quantities~\eq{eq_suscept} and \eq{eq_binder} characterize the fluctuations of the monopole density $\rho$.

The monopole density diminishes with the increase of the coupling constant $\beta$. The position of the deconfining phase transition,
\beqn
\beta_c = 1.010(1), \qquad \mbox{(no plates)}\,,
\label{eq_beta:c:inf}
\eeqn
is well visible from the discontinuity in the monopole density, Fig.~\ref{fig:mdsusc_nocas}(a). The discontinuity -- which is pertinent to the first-order phase transition -- coincides with high accuracy with the positions in the peaks of the susceptibility of the monopole density, Fig.~\ref{fig:mdsusc_nocas}(b) and the corresponding Binder cumulant, Fig.~\ref{fig:mdsusc_nocas}(c). In the vicinity of the phase transition, we take finely spaced values of $\beta$ with the step $\delta \beta = 0.001$ to achieve reasonable accuracy of our simulations in evaluation of the (pseudo)critical values of $\beta_c$ calculated as a maximal value of the derivative of the monopole density with respect to $\beta$, and as a position of the peaks of the susceptibility and the Binder cumulant of monopole density. The position of the pseudocritical point~\eq{eq_beta:c:inf} coincides, within the error bars, with the infinite-volume result of Ref.~\cite{Arnold_2002jk}.

\subsection{Monopole properties with Casimir plates}

Perfectly conducting Casimir plates are introduced via the inhomogeneous coupling~\eq{eq_beta:P} which serves as the Lagrange multiplier that reduces physical fluctuations of the gauge field at the plates. In practice, we take the sufficiently large value of the dielectric constant, $\varepsilon = 10^3$, which corresponds to the asymptotically large coupling constant at the plates, $\beta_P \to \infty$. The coupling constant in bulk (outside and inside the plates) is fixed to take a homogeneous value~$\beta$. We consider the separations $R \equiv |l_1 - l_2| =1,\dots,8$ between the plates. 

\begin{figure}[!htb]
    \centering
\begin{tabular}{cc}
	        \includegraphics[width=0.225\textwidth]{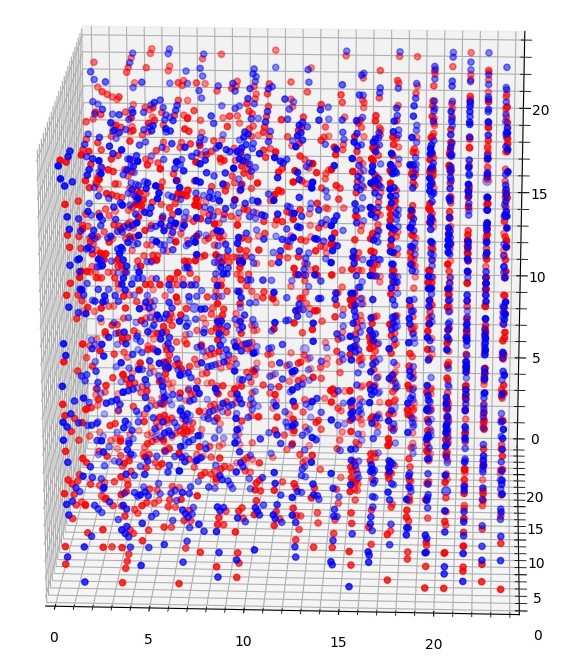} &
	        \includegraphics[width=0.225\textwidth]{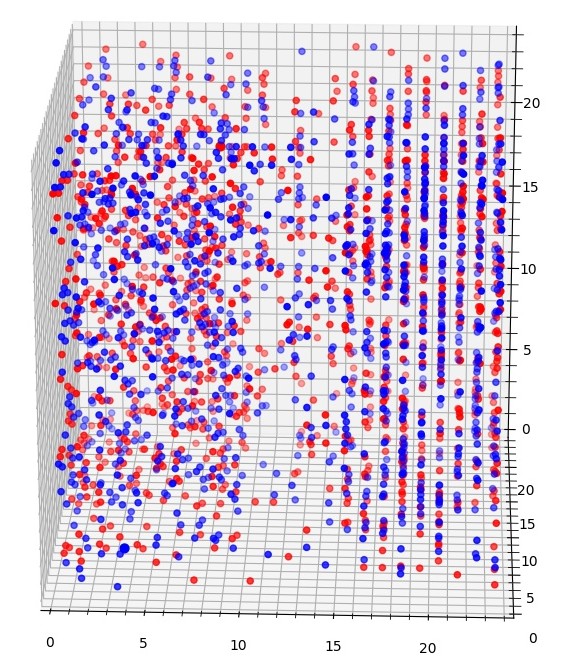} \\
	        (a) & (b)
\end{tabular}
	\caption{Typical examples of monopole configurations in (a) the confining phase ($\beta=0.8$) and (b) the deconfining phase ($\beta = 0.9$) for the plates separated by the distance $R=3$. The monopoles and anti-monopoles are represented by the red and blue dots, respectively. The plates, positioned vertically in the middle of the lattice, are not shown.}
	\label{fig:monoconf}
\end{figure}

We immediately notice that closely separated plates affect the monopoles between them. The effect is readily visible in the examples of the typical monopole configurations shown in Fig.~\ref{fig:monoconf} for two values of the bulk coupling constant $\beta$. The plates tend to diminish the monopole density in the volume between them compared to the monopole density outside the plates. The suppression effect is enhanced for larger values of the coupling constant $\beta$ (at weaker coupling), as one can see from comparisons of Figs.~\ref{fig:monoconf}(a) and (b). The suppression of the monopole density suggests that the confining property should be weakened in between the plates, and therefore the confinement-deconfinement phase transition should occur at stronger values of the coupling constant (smaller $\beta$'s). This observation will be confirmed below.

\begin{figure}[ht]
\centering
  \includegraphics[width=1\linewidth]{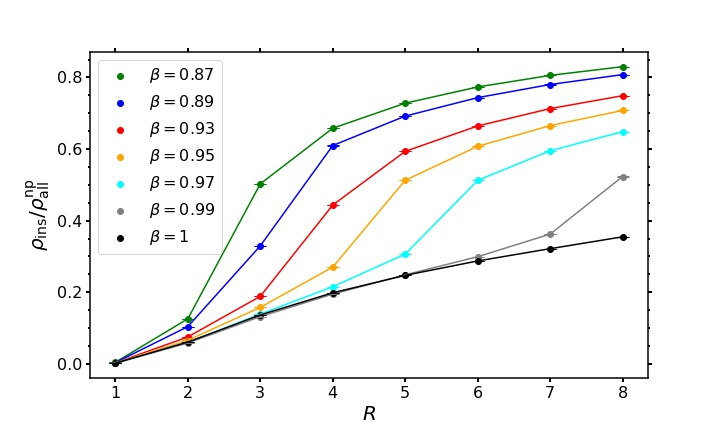}
  \caption{The ratio $\rho_{\mathrm{ins}}/\rho^{\mathrm{np}}_{\mathrm{ins}}$ of the monopole density $\rho_{\mathrm{ins}}$ inside the Casimir plates to the monopole density in the absence of the plates, $\rho_{\mathrm{ins}}^{\mathrm{np}}$, vs. the inter-plate separation $R$ for a fixed set of the lattice coupling $\beta$.}
  \label{fig:monodensnorm}
\end{figure}

The ratio of the monopole density in between the plates $\rho_{\mathrm{ins}}$ and the monopole density at the same $\beta$ in the absence of the plates, $\rho^{\mathrm{np}}_{\mathrm{all}}$, are shown in Fig.~\ref{fig:monodensnorm} for various separations~$R$. All couplings $\beta$ shown in the plot, the shrinking plates lead to the diminishing monopole density. At weaker coupling (larger $\beta$), the relative monopole density is affected stronger than at stronger coupling (smaller $\beta$). The relative monopole density, $\rho_{\mathrm{ins}}/\rho^{\mathrm{np}}_{\mathrm{ins}}$, has an inflection point at certain $R=R^*$ at fixed $\beta$. This point moves towards smaller values of $R$ as the lattice coupling $\beta$ decreases. The latter fact indicates that the model may have a $\beta$-dependent transition which moves towards smaller $R$ as the coupling $\beta$ gets larger. 

\begin{figure*}[!htb]
    \centering
\begin{tabular}{ccc}
	        \includegraphics[width=0.33\textwidth]{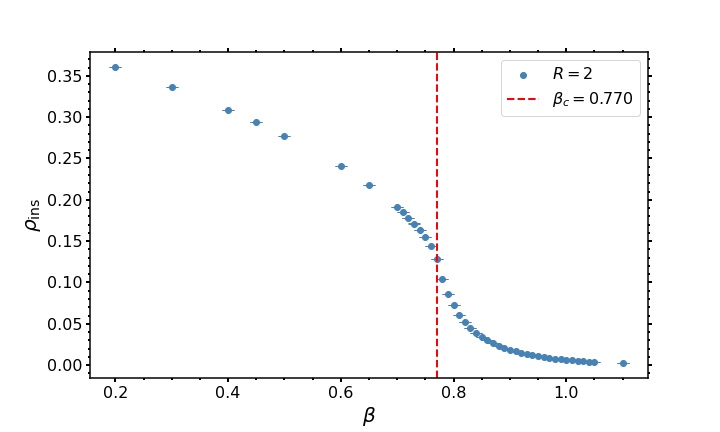}
	    &
	        \includegraphics[width=0.33\textwidth]{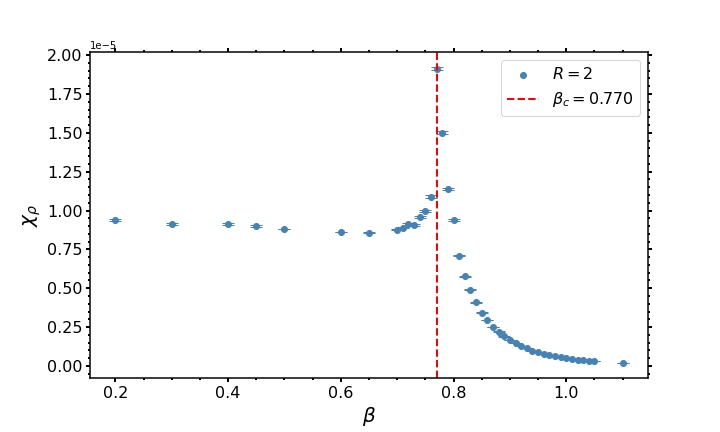}
	    &
	        \includegraphics[width=0.33\textwidth]{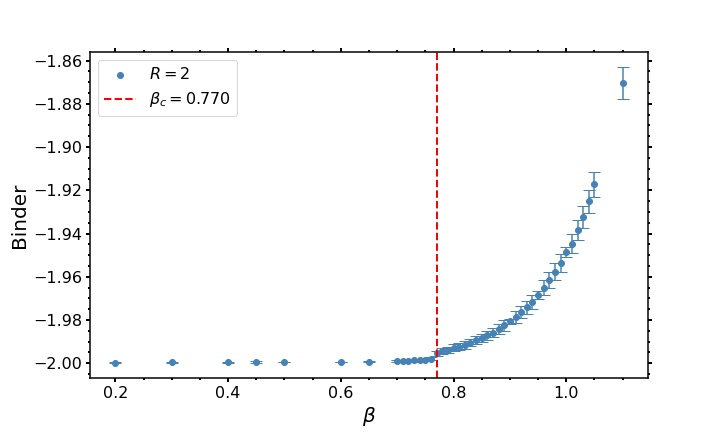}
	     \\
	        \includegraphics[width=0.33\textwidth]{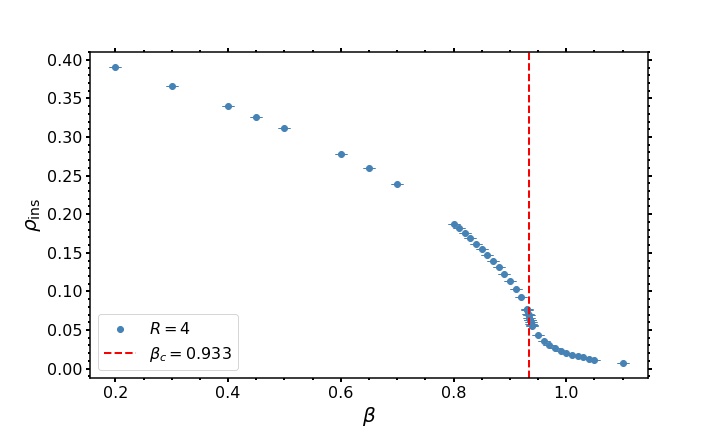}
	    &
	        \includegraphics[width=0.33\textwidth]{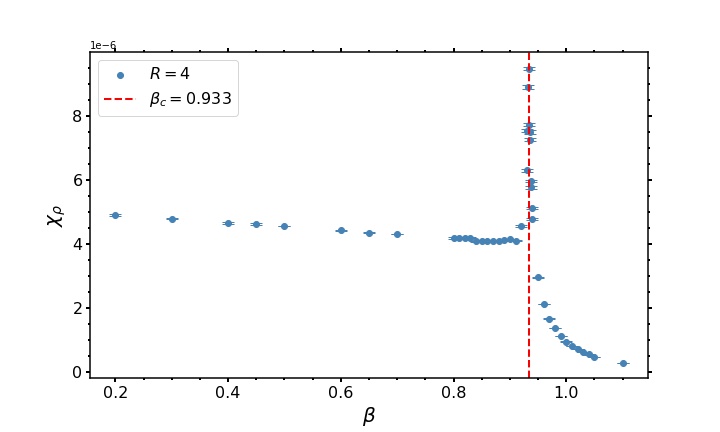}
	    &
	        \includegraphics[width=0.33\textwidth]{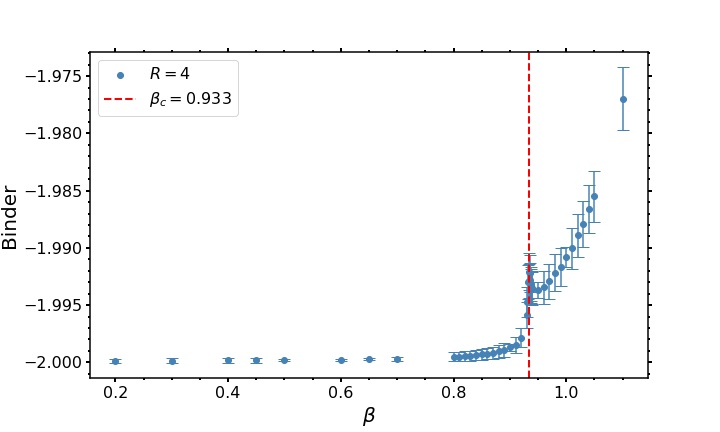}
	    \\
	        \includegraphics[width=0.33\textwidth]{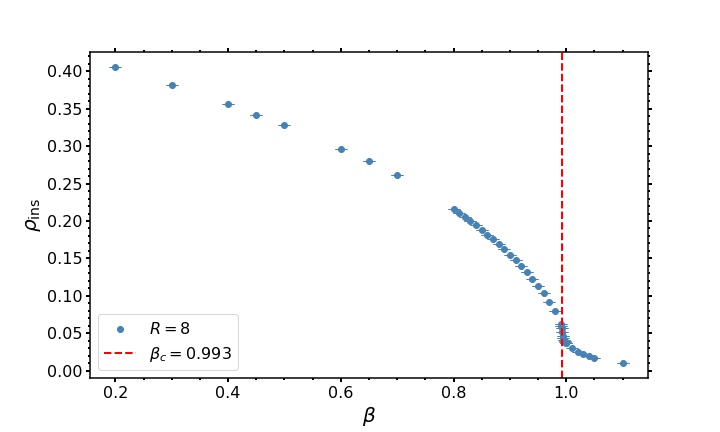}
	    &
	        \includegraphics[width=0.33\textwidth]{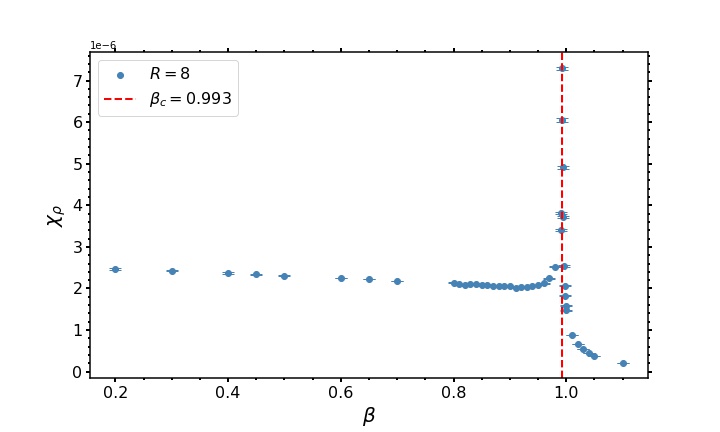}
	    &
	        \includegraphics[width=0.33\textwidth]{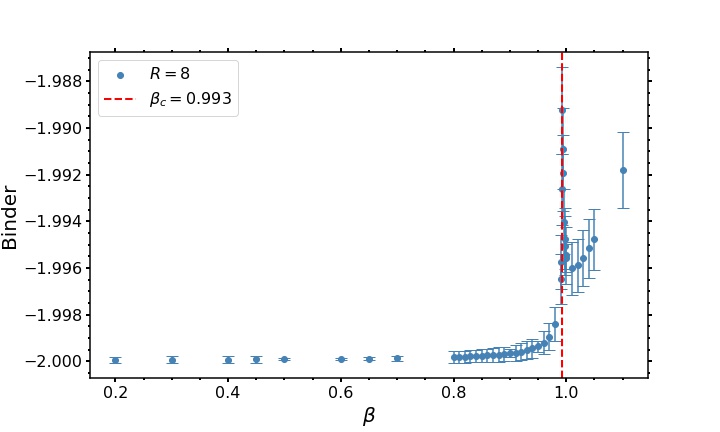}
\end{tabular}
	\caption{The monopole density (the left panel), its susceptibility (the middle panel) and the Binder cumulant (the right panel) plotted as the function of $\beta$ for three separations between the plates (from top to bottom): $R={2,4,8}$. The corresponding critical coupling constants, $\beta_c = \beta_c(R) $, are shown in the insets.}
	\label{fig:mdsuscbinder}
\end{figure*}

In Fig.~\ref{fig:mdsuscbinder}  we show the monopole density in between the plates $\rho_{\mathrm{ins}}$, its susceptibility, and the corresponding Binder cumulant for three values of the inter-plate distance~$R$. One can immediately make a few qualitative observations from these figures. 

Firstly, we notice that all these quantities behave similarly to the case without plates (Fig.~\ref{fig:mdsusc_nocas}) implying affinity of these transitions. Secondly, we notice that for any fixed $R$, the monopole density, its susceptibility, and the Binder cumulant experience the singularities at the same value of the coupling constant~$\beta$, highlighting the presence of a genuine thermodynamic instability. Thirdly, the positions of these singularities, $\beta_c = \beta_c(R)$, shift towards the strong coupling region as the distance between $R$ the plates diminishes. In other words, the closer the plates, the weaker the monopole component of the vacuum. 

\begin{figure}[ht]
\centering
  \includegraphics[width=1.05\linewidth]{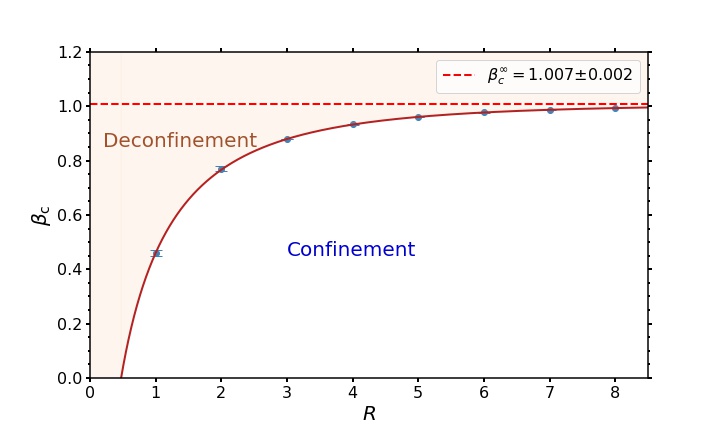}
  \caption{The phase diagram of the vacuum of the compact U(1) gauge theory in between the perfectly metallic plates separated by the distance $R$. The solid line represents the best fit~\eq{eq_fit} of the critical coupling $\beta_c$ which corresponds to the confinement-deconfinement phase transition at the inter-plate distance~$R$. The $R \to \infty$ limit is shown by the dashed horizontal line.}
    \label{fig:phase:diagram}
\end{figure}

The dependence of the critical coupling $\beta_c$ of the inter-plate distance $R$ is shown in Fig.~\ref{fig:phase:diagram}. As we discussed above, the smaller $R$ the smaller $\beta_c$. The dependence can be fitted by the function 
\beqn
\beta_c^{\mathrm{fit}}(R) = \beta^{\infty}_c - \alpha \exp{\left[-(R^2/R_0^2)^\nu\right]}\,, 
\label{eq_fit}
\eeqn
where the best fit parameter $\beta_c = 1.0071(16)$ gives us the critical coupling at the infinitely separated plates. This value is close to the transition point in the absence of the plates~\eq{eq_beta:c:inf}, highlighting the self-consistency of our approach. The other best fit parameters in Eq.~\eq{eq_fit} are as follows: $\alpha =  3.7(6)$, $R_0 = 0.28(7)$ and $\nu = 0.257(16)$. The best fit function is shown in Fig.~\ref{fig:phase:diagram} by the solid line.

The critical coupling~\eq{eq_fit} separates the confinement phase ($\beta < \beta_c$) and the deconfinement phase ($\beta > \beta_c$). The phase transition line is a rising function of the inter-plate distance $R$. The critical coupling vanishes, $\beta_c^{\mathrm{fit}}(R_c) = 0$, at the critical distance $R_c = 0.47(7)$. Formally, at the separations smaller than critical value, $R < R_c$, the theory cannot reside in the confining phase. Since the inter-plate separation is a positive integer number in the discretized theory, $R = 1,2,\dots$, this asymptotic critical value cannot be reached on the lattice. 

The loss of the confinement property of the vacuum between the metallic plates in 3+1 dimensions, revealed by Fig.~\ref{fig:phase:diagram}, can be understood using the analogy with the similar effect in the 2+1 dimensional theory~\cite{ref_paper_2}. Namely, the monopoles and anti-monopoles interact with each other via the long-range massless photon exchange. In the absence of the plates, this interaction, in 3+1 dimensions, decays as $|x|^{-2}$ as the 4-dimensional distance $x$ between the objects increases. In the presence of the plates, the system experiences the dimensional reduction from four- to three-dimensional spacetime. In the latter, the interaction between the (anti-)monopoles strengthens and decays slower, as $|x|^{-1}$. These two factors lead to the breaking of the infrared monopole clusters into smaller clusters and, consequently, to the disappearance of the monopole condensate.

In the 2+1 dimensional model, the same effect leads to the pairing of monopoles and anti-monopoles into the magnetically neutral monopole pairs (the lower-dimensional counterparts of the small clusters) and to the decay of the Coulomb monopole gas (the lower-dimensional analogue of the monopole condensate). The neutral pairs (the small clusters) cannot support the confinement and confining property is lost between the sufficiently close plates~\cite{ref_paper_2,ref_paper_3}.

\section{Deconfinement order parameter}
\label{sec_confinement}

We determined the nature of confinement and deconfinement phases in the whole phase diagram of Fig.~\ref{fig:phase:diagram} using the simple fact that these phases at finite separation $R$ are smoothly connected to the known phases in the $R \to \infty$ limit. In this section, we quantify this assertion by calculating the deconfinement order parameter, the expectation of the Polyakov loop, in between the plates.

Usually, the Polyakov loop is determined at a finite temperature where the extension of the lattice in the imaginary time direction is finite. The same order parameter can also be used on the zero-temperature lattice with a finite extension in the temporal direction, $N_T$. 

In the Abelian gauge theory, the Polyakov loop $P_{\bs x}$ at the spatial space point ${\bs x}$ is given by a cyclic product of temporal link variables:
\begin{equation}
    P_{\bs x} = \prod_{x_4=0}^{N_T - 1} e^{i \theta_{{\bs x},x_4;\mu = 4}}\,.
\label{eq_ploop}
\end{equation}
Due to the property of the temporal cyclicity, this quantity does not depend on the time slice where it is defined.

The expectation value of this gauge-invariant quantity, $P = \left\langle P_{\bs x}\right\rangle$ serves as a deconfinement order parameter: at an infinite-volume lattice, $P \neq 0$ in the deconfinement phase and $P=0$ in the confinement phase. At finite lattice (as in our case), the expectation value of the Polyakov loop is nonzero in both phases, being small (large) in the confinement (deconfinement) phase.

We compute the expectation value of the modulus of the operator~\eq{eq_ploop} averaged over a set of lattice points in a fixed timeslice of the 3-dimensional volume V${}_3$:
\begin{equation}
    |P| = \left|\frac{1}{V_3}\sum\limits_{{\bs x} \in V_3}P_{\bs x}\right|\,.
\label{eq_ploopmod}
\end{equation}

\begin{figure}[!htb]
            \centering
            \includegraphics[width=0.45\textwidth]{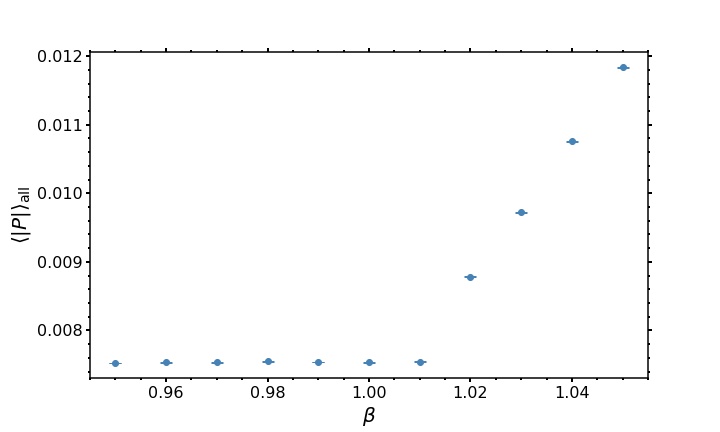}
        \caption{The Polyakov loop as the function of lattice coupling $\beta$ in the absence of plates.}
\label{fig:ploop_nocas}
\end{figure}

The expectation value of the Polyakov order parameter~\eq{eq_ploopmod} in the absence of the plate is shown in Fig.~\ref{fig:ploop_nocas}. We show the region around the phase transition point~\eq{eq_beta:c:inf}, where the change in the behavior is well seen.

\begin{figure}[!htb]
	\centering
	\includegraphics[width=0.45\textwidth]{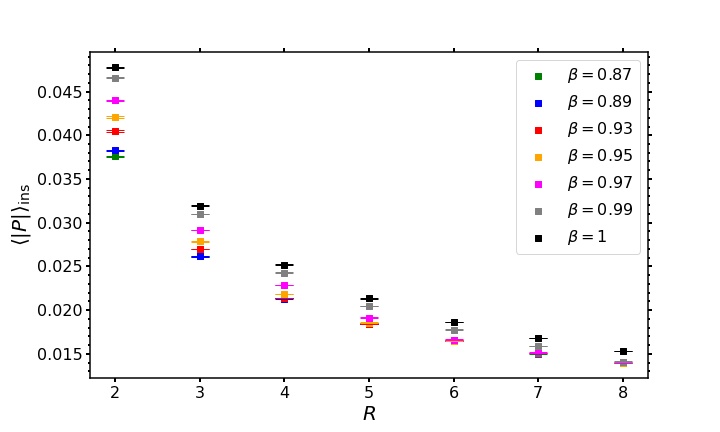}
	\caption{The expectation value of the Polyakov loop in the space between the Casimir plates at the separation $R$ at a set of fixed coupling constants $\beta$.} 
	\label{fig:ploop_nocas_diffb}
\end{figure}

\begin{figure}[!htb]
	\centering
	\includegraphics[width=0.45\textwidth]{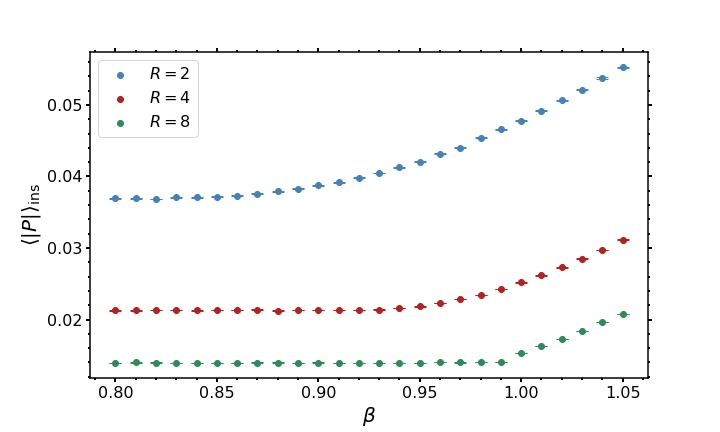}
	\caption{\small{The Polyakov loop inside the plates vs $\beta$ at fixed~$R$.}} 
	\label{fig:ploop}
\end{figure}

The order parameter evaluated in the space between the plates is presented in Fig.~\ref{fig:ploop_nocas_diffb}. We show this quantity, as a function of the inter-plate separation $R$, for the same set of $\beta$ values as used in Fig.~\ref{fig:monodensnorm} for the inter-plate monopole density. The shrinking plates induce the deconfining phase resulting in the increase of the Polyakov loop (Fig.~\ref{fig:ploop_nocas_diffb}) in agreement with the diminishing monopole density (shown in Fig.~\ref{fig:monodensnorm}). The effect appears to work at all values of the coupling constant $\beta$. The similar tendency is seen in Fig.~\ref{fig:ploop} which shows the same quantity vs. $\beta$ at a set of fixed inter-plate separations~$R$.

These results support the phase diagram of Fig.~\ref{fig:phase:diagram}.

\section{Conclusions}

Using the first-principle numerical simulations, we show that the vacuum of confining gauge theory, the compact U(1) gauge model in 3+1 dimensions, is affected by closely spaced perfectly conducting parallel plates. The non-perturbative Casimir effect modifies the vacuum structure and leads to the Casimir-induced deconfining phase transition in between the plates. Our main result, the phase diagram in the plane ``coupling constant''--``distance between the plates'' is given in Fig.~\ref{fig:phase:diagram}. 

\acknowledgments
The numerical calculations were performed in 2021 at the computing cluster Vostok-1 of the Far Eastern Federal University. The work of VAG, AVM, and AST was supported by Grant No. 0657-2020-0015 of the Ministry of Science and Higher Education of Russia.

%


\end{document}